\documentclass[11pt,twoside]{article}
\usepackage{asp2010}
\usepackage{graphicx}

\resetcounters

\begin{document}

\title{Five and a half roads to form a millisecond pulsar}
\author{Thomas~M.~Tauris$^{1,2}$
\affil{$^1$Argelander-Insitut f\"ur Astronomie, Universit\"at Bonn, Auf
  dem H\"ugel 71, 53121 Bonn, Germany}
\affil{$^2$Max-Planck-Institut f\"ur Radioastronomie, Auf dem H\"ugel 69,
  53121 Bonn, Germany}}

\begin{abstract}
In this review I discuss the characteristics and the formation of all classes of millisecond pulsars (MSPs).
The main focus is on the stellar astrophysics of X-ray binaries
leading to the production of fully recycled MSPs with white dwarf (WD) or substellar semi-degenerate companions. 
Depending on the nature of the companion star MSPs are believed to form
from either low-mass {X}-ray binaries (LMXBs) or intermediate-mass {X}-ray binaries (IMXBs).
For each of these two classes of {X}-ray binaries
the evolutionary status of the donor star
-- or equivalently, the orbital period -- at the onset of the Roche-lobe overflow (RLO) is the determining factor for 
the outcome of the mass-transfer phase and thus the nature of the MSP formed.
Furthermore, the formation of binary MSPs is discussed in context of the $P$$\dot{P}$-diagram, as well as new interpretations
of the Corbet diagram.
Finally, I present new models of Case~A RLO of IMXBs in order to reproduce the two~solar~mass pulsar PSR J1614$-$2230.
\end{abstract}

\section{Introduction}
MSPs are known to be important sources of research in many areas of fundamental physics.
Equally important, however,
binary MSPs represent the end point of stellar evolution, and their observed orbital
and stellar properties are fossil records of their evolutionary history. Thus one
can also use binary pulsar systems as key probes of stellar astrophysics.
It is well established that the neutron star in binary MSP systems forms first, descending from the
initially more massive of the two binary stellar components. The neutron star is subsequently spun-up to a high spin frequency
via accretion of mass and angular momentum once the secondary star evolves
\citep{acrs82,rs82,bv91}. During this recycling phase the
system is observable as an X-ray binary \citep[e.g.][]{nag89,bcc+97} and towards the end of this phase
as an {X}-ray millisecond pulsar \citep{wv98,asr+09}.
Although this formation scenario is now commonly accepted,
many aspects of the mass-transfer process and the accretion physics (e.g. details of non-conservative evolution
and the accretion efficiency) are
still not well understood. Some of these issues will be addressed here. Given the limited number of pages
in this conference proceedings review the discussion will mainly be of a qualitative character. 
For a more detailed review, see e.g. \citet{bv91,tv06}.

\section{Mass transfer in {X}-ray binaries and the nature of the donor star}
\begin{figure*}[h]
\begin{center}
\mbox{\includegraphics[width=0.43\textwidth, trim = 50 50 50 50, angle=0]{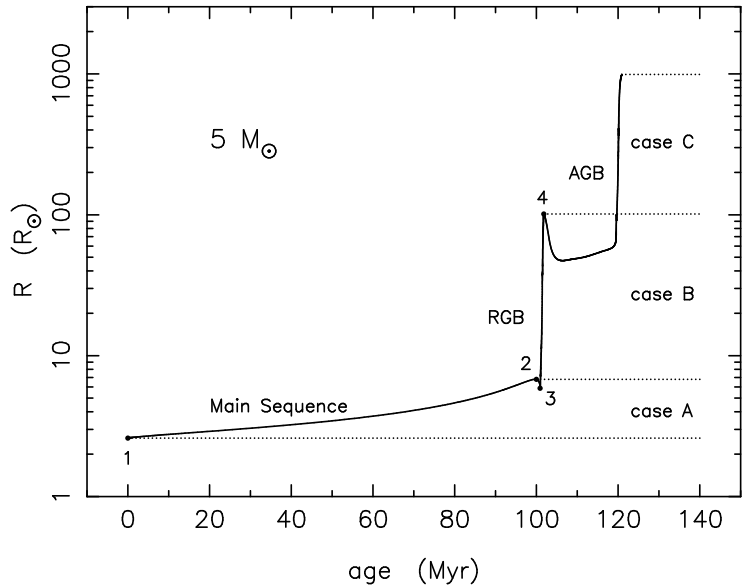}}
  \caption[]{
    Stellar radius as a function of age for a  $5\,M_{\odot}$ star. Also shown is the definition of RLO Cases A, B and C.
    Figure from \citet{tv06}.
    }
\label{fig:RLOcases}
\end{center}
\end{figure*}

Consider a close interacting binary system which consists of a non-degenerate donor star and a compact object, in our case a neutron star.
If the orbital separation is small enough the (evolved) non-degenerate star fills its inner common equipotential
surface (Roche-lobe) and becomes a donor star for a subsequent epoch of mass transfer toward the, now, accreting neutron star.
In this phase the system is observed as an {X}-ray binary.
When the donor star fills its Roche-lobe it is perturbed by removal of mass and it falls out of hydrostatic
and thermal equilibrium. In the process of re-establishing equilibrium the star will either grow or shrink
-- depending on the properties of its envelope layers -- first on a dynamical (adiabatic)
timescale and subsequently on a slower thermal timescale. However,
any exchange and loss of mass in such an {X}-ray binary system will also lead to alterations of the orbital dynamics
via modifications in the orbital angular momentum,
and hence changes in the size of the critical Roche-lobe radius of the donor star. The stability of the
mass-transfer process therefore depends on how these two radii evolve (i.e. the radius of the star and the Roche-lobe radius).
The various possible
modes of mass exchange and loss
include, for example, direct fast wind mass loss, Roche-lobe overflow (with or without isotropic re-emission)
and common envelope evolution \citep[e.g.][and references therein]{vdh94a,spv97}.
The RLO mass transfer can be initiated while the donor star is still on the main sequence (Case~A RLO), during hydrogen
shell burning (Case~B RLO) or during helium shell burning (Case~C RLO) -- see Fig.~\ref{fig:RLOcases}.
The corresponding evolutionary timescales for these different cases will in general proceed on
a nuclear, thermal or dynamical timescale, respectively, or a combination thereof. This timescale is important for
the extent to which an old neutron star can be recycled (i.e. with respect to its final spin period and {B}-field).

The dynamical evolution of a binary system can be found by solving for the changes in the orbital separation, $a$.
The orbital angular momentum of a circular binary system is given by:
$J_{\rm orb} = \mu\,\Omega \,a^2$,
where $\mu$ is the reduced mass 
and the orbital angular velocity is:
$\Omega = \sqrt{GM/a^3}$.
A simple logarithmic differentiation of the orbital angular momentum equation yields the rate of change in orbital separation:
\begin{equation}
  \frac{\dot{a}}{a} = 2\,\frac{\dot{J}_{\rm orb}}{J_{\rm orb}} - 2\,\frac{\dot{M}_1}{M_1} 
                    - 2\,\frac{\dot{M}_2}{M_2} + \frac{\dot{M}_1 + \dot{M}_2}{M}
  \label{adot}
\end{equation}
where the two stellar masses are given by $M_1$ and $M_2$, and the total change in orbital angular momentum per unit time is given by: 
$\dot{J}_{\rm orb}=\dot{J}_{\rm gwr}+\dot{J}_{\rm mb}+\dot{J}_{\rm ls}+\dot{J}_{\rm ml}$.
These four terms represent gravitational wave radiation, magnetic braking, spin-orbit couplings and mass loss, respectively
\citep[e.g.][and references therein]{tv06}.

\begin{figure*}[ht!]
\begin{center}
\mbox{\includegraphics[width=0.97\textwidth, angle=0]{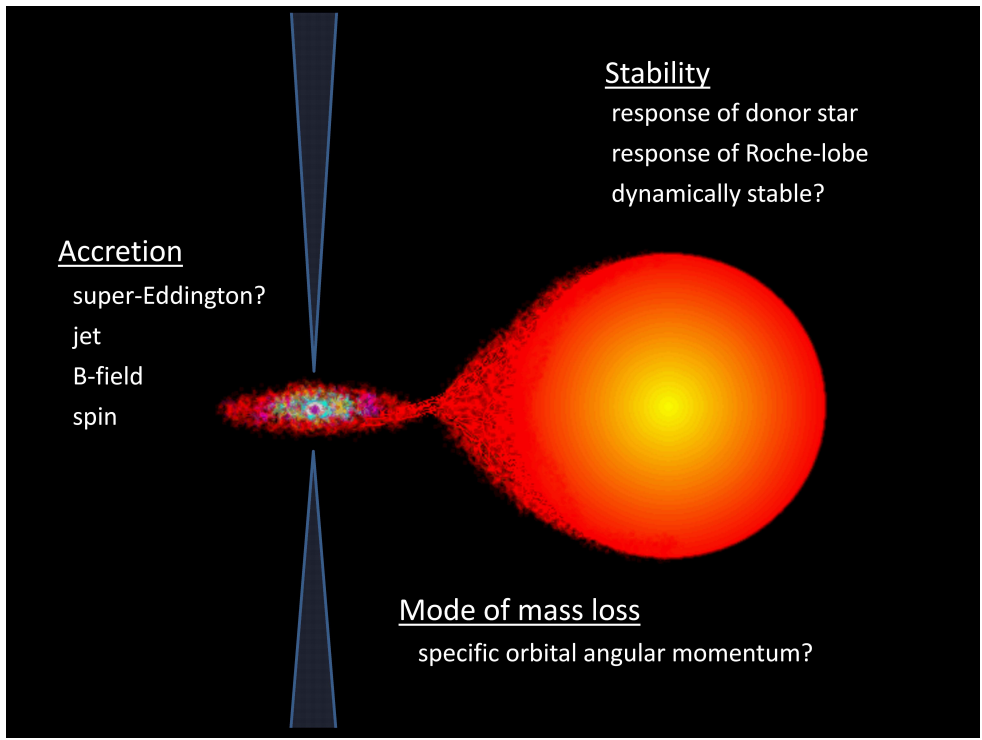}}
  \caption[]{Illustration of some of the many issues to consider in {X}-ray binary models.
    }
\label{fig:Tauris_X-1}
\end{center}
\end{figure*}

When modelling the evolution of an {X}-ray binary one should take into account a number of issues related to: 
the stellar evolution, the stability of the mass-transfer process, the ejection of matter from the system and the accretion of material onto
the neutron star -- see illustration in Fig.~\ref{fig:Tauris_X-1}. Ideally, all these calculations should be performed self-consistently.
The major uncertainties here are related to the amount and the mode of the specific orbital angular momentum of ejected matter -- and for
close systems, also the treatment of the spin-orbit couplings.

\begin{figure*}[ht!]
\begin{center}
\mbox{\includegraphics[width=0.70\textwidth, angle=-90]{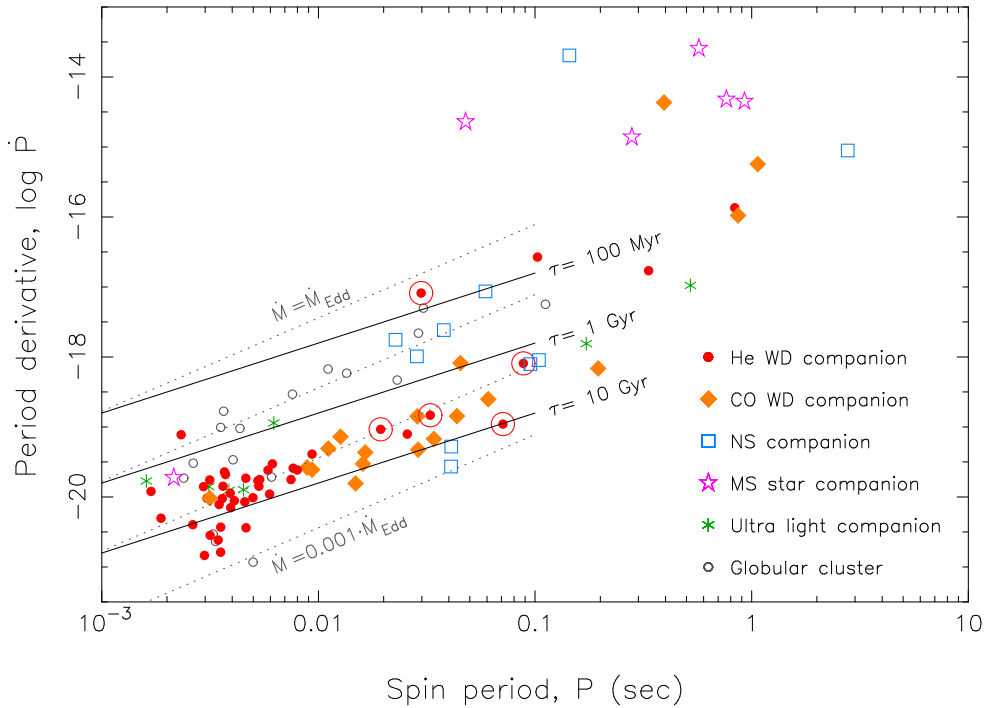}}
  \caption[]{
    The distribution of the $\sim\!100$ binary radio pulsars with measured values in the $P\dot{P}$-diagram.
    The symbols to the right indicate the nature of the companion star. Pulsars with a He~WD companion marked by a ring
    are discussed in Section~\ref{sec:corbet}. Data taken from the {\it ATNF Pulsar Catalogue} in May 2011
    \citep[][http://www.atnf.csiro.au/research/pulsar/psrcat]{mhth05}.
    }
\label{fig:ppdot}
\end{center}
\end{figure*}

\section{The population of binary millisecond pulsars}
Fig.~\ref{fig:ppdot} shows the location of the known binary pulsars in the $P\dot{P}$-diagram. 
The fully recycled MSPs ($P<10\,{\rm ms}$) are dominated by having mainly He~WD companions,
although also ultra light (substellar) semi-degenerate companions are seen  -- often in eclipsing "black widow-like systems" --
as well as a few systems with the more massive CO~WD companions.
The mildly recycled MSPs ($10\,{\rm ms} < P < 100\,{\rm ms}$) are dominated by CO~WD (or ONeMg~WD) and neutron star companions.
As we shall see, this relatively slow spin rate is expected from an evolutionary point of view as a consequence of the rapid mass-transfer phase
from a relatively massive donor star.
The pulsars with similar spin periods and He~WD companions apparently also had a limited recycling which may hint their origin
-- see Section~\ref{sec:corbet}. 
The double neutron star systems descend from high-mass X-ray binaries (HMXBs) and are not discussed in 
this review \citep[see e.g.][and references therein]{vt03,dp03,dps06}.\\
MSPs are not only characterized by a rapid spin. All of them also possess a low surface magnetic flux density, $B$
which is typically of the order $10^8\,{\rm G}$, or some 3-5 orders of magnitudes less than the B-fields of ordinary, non-recycled pulsars
(see discussion in Section~\ref{sec:spinup}).
Hence, the recycled pulsars do not suffer as much from loss of rotational energy due to emission of magnetodipole waves.
For this reason the MSPs have small period derivatives $\dot{P}<10^{-18}$ and hence they are able to maintain the 
production of radio waves keeping them observable for billions of years.\\ 
Binary pulsars in globular clusters are, in general, not suitable as tracers of their stellar evolution history because
of the frequent encounters and
exchanges of companion stars in the dense environments \citep{rhs+05}. Hence, this review only considers binary pulsars formed in the Galactic disk.

\begin{figure*}[h!]
\begin{center}
\mbox{\includegraphics[width=0.70\textwidth, angle=-90]{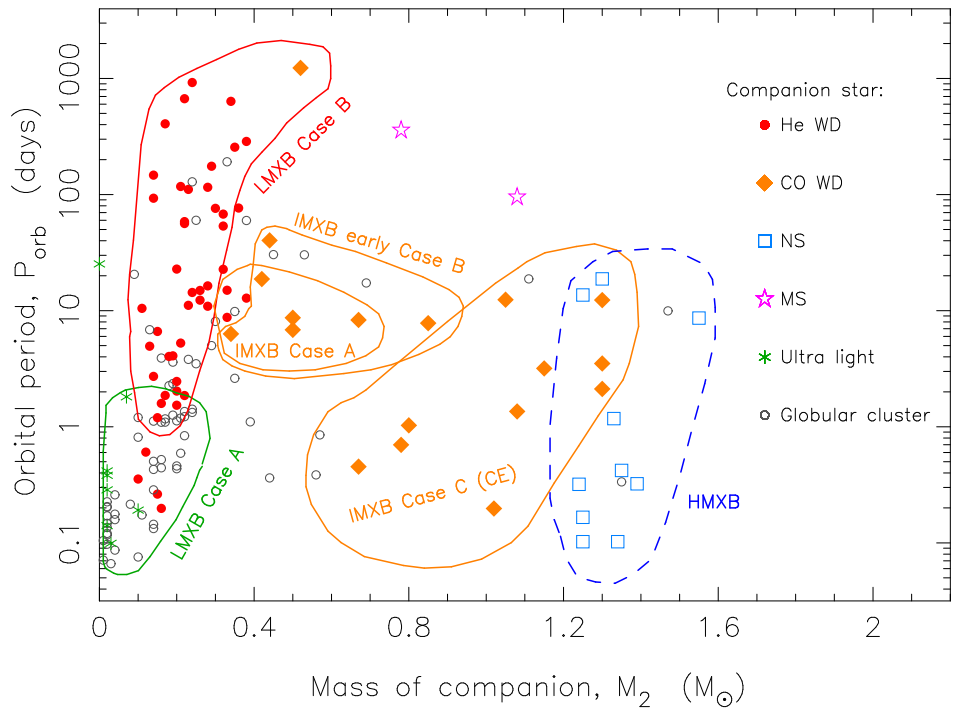}}
  \caption[]{
    The distribution of 162 binary radio pulsars in the companion mass-orbital period plane.
    Data taken from the {\it ATNF Pulsar Catalogue} in May 2011.
    The five roads to form an MSP are shown: Two roads from LMXBs (Cases A and B) and three roads
    from IMXBs (Cases A, B and C), see also Fig.~\ref{fig:cartoon} and Table~\ref{table:classes}.
    HMXBs leave behind the double neutron star systems and only partially
    recycle the first born neutron star (see their location in Fig.~\ref{fig:ppdot}). The boundaries are not very strict and keep
    in mind that most of the companion masses have large error bars.
    }
\label{fig:MPrel_all}
\end{center}
\end{figure*}

When discussing the origin of MSPs with various companion stars it makes sense to plot the binary orbital period
as a function of the companion star mass -- see Fig.~\ref{fig:MPrel_all}. This figure is essential for
understanding the progenitor systems.

\section{MSPs with He~WD or substellar dwarf companions -- evolution from LMXBs}
To the left in Fig.~\ref{fig:MPrel_all} one sees the MSPs with either ultra light companions or
He~WD companions. Having low-mass companions these systems thus descend from LMXBs. 
MSPs with $P_{\rm orb} \le 1\,{\rm day}$ originate from LMXBs in very tight orbits
where the donor star already initiated RLO while it was still on the main sequence (thus Case~A RLO), whereas the
MSPs with $P_{\rm orb} > 1\,{\rm day}$ originate from wider orbit LMXBs where the donor star did not fill its Roche-lobe
until it had evolved and expanded to become a (sub)giant, i.e. Case~B RLO.
It has been shown by \citet{ps88,ps89} that a critical orbital bifurcation period ($P_{\rm bif}$) exists
at the onset of the RLO, separating the formation of {\it converging} LMXBs from {\it diverging} LMXBs.
The theoretical estimated value of $P_{\rm bif}$ is $\sim\!1\,{\rm day}$, 
but depends strongly on the treatment of tidal interactions and the assumed strength of magnetic braking which drains
the system of orbital angular momentum \citep{vvp05,ml09}.

\subsection{Close LMXB systems -- formation of black-widow pulsars}
In observed MSP systems with $P_{\rm orb} \le 1\,{\rm day}$ the mass transfer was driven by loss of orbital angular momentum due to magnetic braking and (for
more narrow systems) emission of gravitational waves. The donor star becomes degenerate, or semi-degenerate, as the ultra-compact binary evolves.
An additional mass-transfer phase is initiated in those final NS+WD systems where emission of gravitational waves drives the two stellar components together. 
The outcome is ultra-compact systems (e.g. {X}-ray transient accreting MSPs) with orbital periods less than one hour and $M_2\simeq 10^{-2}\;M_{\odot}$
\citep{db03}.
Furthermore, irradiation effects, first by the {X}-ray flux from the accreting neutron star \citep{pod91} and later by the pulsar wind \citep{tav92},
as well as tidal dissipation of energy in the envelope \citep{as94}, may cause the companion star to be thermally bloated and evaporate. 
In some cases this combined evolution eventually leads to the formation of pulsar planets \citep[PSR~B1257+12,][]{wf92} and solitary MSPs \citep[PSR~B1937+21,][]{bkh+82}.
Many radio pulsars in narrow systems display eclipses during their orbital motion and in the case of PSR~J2051$-$0827 one can
even measure the effects of gravitational quadrupole moment changes \citep{lvt+11}.
In those binaries where the companion leaves behind a low-mass He~WD ($< 0.20\;M_{\odot}$) the cooling age determination
of the WD (and thus the age of the MSP) is complicated by their thick residual hydrogen envelopes.
Pycno-nuclear burning at the bottom of these envelopes can keep the low-mass He-WDs warm for $\sim\!10^9\;{\rm yr}$ \citep[see discussion in][]{vbjj05}.

\subsection{Wide LMXB systems -- formation of classic MSPs with He~WD companions}
\begin{figure*}[h]
\begin{center}
\mbox{\includegraphics[width=0.70\textwidth, angle=-90]{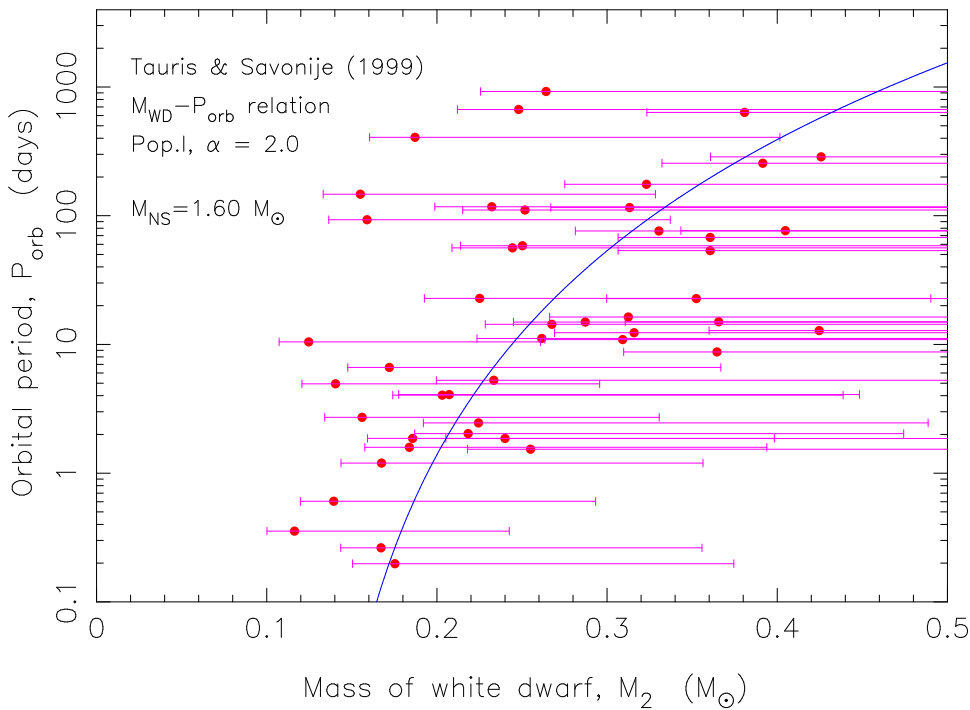}}
  \caption[]{
    Orbital period as a function of WD mass for MSPs with He~WD companions. The error bars
    of the WD masses represent the unknown orbital inclination angle -- the left end corresponds
    to an inclination angle of $90\deg$ and the right end marks the 90\% probability limit.
    The additional assumption is that all neutron stars are assumed to have a mass of $1.60\,M_{\odot}$ (for simplicity).
    This mass is larger than the typical $\sim\!1.35\;M_{\odot}$ found for pulsars
    in double neutron star systems. However, fully recycled MSPs are expected to have accreted significantly
    more material.
    The theoretical ($M_{\rm WD},\,P_{\rm orb}$)-correlation is shown as a blue line \citep{ts99}.
    }
\label{fig:MPrel}
\end{center}
\end{figure*}

In wide-orbit LMXBs the donor star did not fill its Roche-lobe until it moved up the red giant branch (RGB).
For low-mass stars ($<2.3\,M_{\odot}$) on the RGB there is a well-known relationship between the mass of
the degenerate helium core and the radius of the giant star -- almost entirely independent of the
mass present in the hydrogen-rich envelope \citep{rw71,wrs83}. 
This relationship is very important for the formation of wide-orbit MSPs 
since it results in a unique relationship between orbital period and
white dwarf mass \citep{sav87,jrl87,rpj+95,ts99}. The companions here are He~WDs with masses $0.2 \le M_{\rm WD}/M_{\odot} < 0.46$ --
unless the system was initially so wide that the donor ascended the asymptotic giant branch (AGB) before initiating RLO and thus left behind
a CO~WD companion (as observed only in PSR~B0820+02 which has an orbital period of 1232~days).
The correlation between $P_{\rm orb}$ and $M_{\rm WD}$ is difficult to verify observationally since very
few MSPs have accurately measured masses of their companion. From the observed mass function
one can only estimate the WD mass which depends on the orbital inclination angle of the system as well
as the neutron star mass -- both of which are unknown. The result of assuming a fixed neutron star mass
which is the same for all systems and plotting the orbital periods as
a function of 
the median He~WD masses (i.e. assuming an inclination angle, $i=60\deg$ for all the systems) can be seen
in Fig.~\ref{fig:MPrel}. 
At first sight this correlation is not obvious under these assumptions --
not even when considering that the correlation depends on the chemical
composition of the donor star as well as the treatment of mixing. 
However, \citet{vbjj05} have shown that the relation is indeed quite good if one only
considers those 7-8 systems for which the WD mass has been estimated fairly accurately. On the other hand, 
there might be a systematic
deviation from the correlation for pulsars with $P_{\rm orb}> 100\;{\rm days}$ \citep{tau96,ts99,sfl+05},
although this claim is based on very small number statistics.
Hopefully, future mass determinations of some of these wide binaries can help to settle this issue.

Finally, it should be mentioned that these wide-orbit binary MSPs also have another fossil of the mass-transfer phase: the correlation
between orbital eccentricity and orbital period. These so-called residual eccentricities are, in general, very small 
(typically between $10^{-6}-10^{-3}$).
The correlation \citep{phi92} is related to tides and arises because 
density fluctuations in the (convective) envelope increase with more evolved donor stars, which have wider orbits, thus
preventing perfect circularization.

\section{MSPs with CO~WD companions -- evolution from IMXBs}
CO~WDs (and ONeMg~WDs) are substantially more massive than He~WDs. Therefore, MSPs with these more massive WDs
must originate from binaries with more massive WD progenitor stars -- i.e. IMXBs which have donor masses of typically
$3-6\;M_{\odot}$. 
Observationally, these systems were first identified by \citet{cam96} as an independent class of MSPs.
There seems to be three roads to produce such systems (see below).

\subsection{Wide IMXB systems -- Case~C RLO and common envelope evolution}
It was demonstrated nicely by \citet{vdh94b} that binary systems with a mildly recycled MSP and 
a CO~WD companion could form from an IMXB with a donor star on the AGB, leading to common envelope (CE)
and spiral-in evolution once the RLO sets in. This model is in particular the favoured scenario for the formation
of very tight binary MSPs  with orbital periods $P_{\rm orb}<3\,{\rm days}$. The reason for this is the
ability of the spiral-in phase to reduce the orbital angular momentum, and thus the orbital period, by a huge amount
\citep[e.g.][and references therein]{dt00}.
One point to notice from this model is that the CE phase is expected to be extremely short.
The in-spiral proceeds on a dynamical timescale (a few orbital periods) followed by an envelope ejection 
phase that may take up to $10^3\,{\rm years}$. However, at least $\sim 0.01\,M_{\odot}$ is needed
to spin up a pulsar to about 10~ms (see Section~\ref{sec:spinup}) and this requires some 0.4~Myr of efficient accretion
at the Eddington limit (a few times $10^{-8}\;M_{\odot}\,{\rm yr}^{-1}$). 
It is therefore believed that the recycling of the MSP may
actually take place from either wind accretion from the naked post-CE core, or from Case~BB RLO
if the naked helium star
expands and fills its Roche-lobe (again) leading to additional recycling of the neutron star.

\subsection{Hertzsprung gap IMXB systems -- early Case~B RLO and isotropic re-emission}
An alternative way of producing MSPs with CO~WD companions was demonstrated by \citet{tvs00}.
They considered donor stars which had just left the main sequence (i.e. Hertzsprung~gap, or sub-giant stars) and applied
the so-called isotropic re-emission model of an IMXB. In this model \citep{bv91}, the far majority
of the matter -- being transfered at a highly super-Eddington rate -- is ejected (e.g. in a jet) 
with the specific orbital angular momentum of the 
accreting neutron star. This model can stabilize the RLO and prevent it from becoming dynamically unstable.
The typical donor star masses in this scenario are $3-6\,M_{\odot}$ and they leave behind a
CO~WD.
Hence, the outcome of this formation scenario is mildly recycled MSPs with CO~WD companions and orbital periods
between 3 and 50~days.
\begin{figure*}[h]
\begin{center}
\mbox{\includegraphics[width=0.70\textwidth, angle=-90]{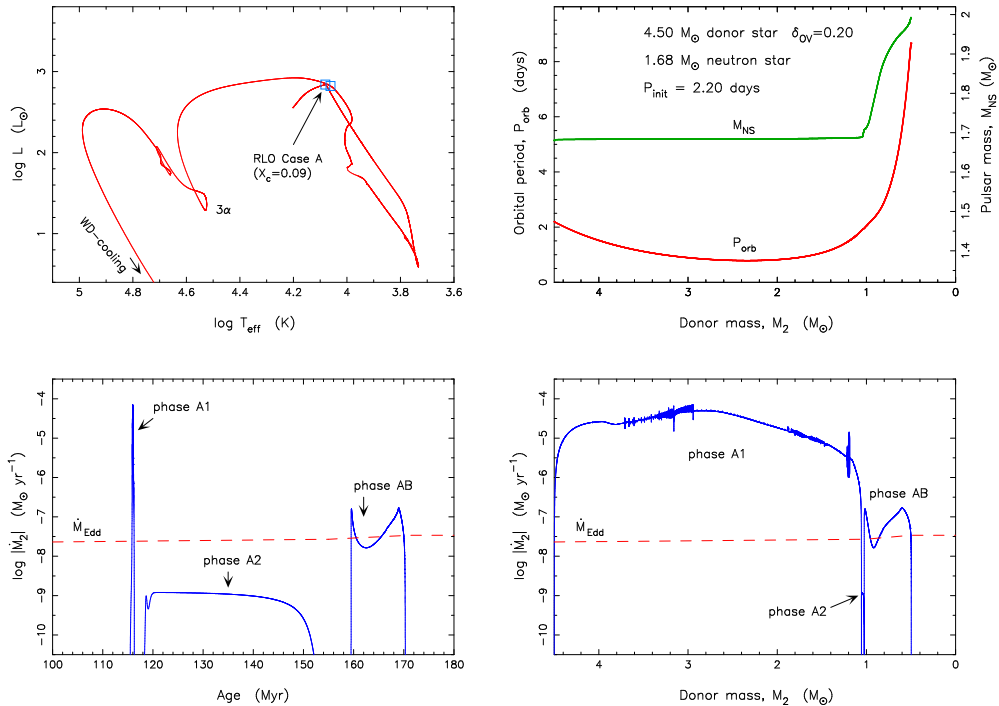}}
  \caption[]{
    The evolution of an IMXB system undergoing Case~A RLO. These models were calculated
    in order to reproduce the record high-mass ($1.97\,M_{\odot}$) MSP J1614$-$2230.
    The mass transfer takes place when the donor star in the HR-diagram (upper left panel) is on the track between the open squares,
    including the loop down to $\log T_{\rm eff}\simeq 3.7$.
    Three phases of mass transfer (A1, A2 and AB) 
    are identified within this track -- see two bottom panels which show the mass-transfer rate 
    as a function of age and donor star mass, respectively. The neutron star is spun~up to an MSP during phase AB
    where it accretes the far majority of its accumulated material (upper right panel).
    We found that this neutron star was born with an initial mass of $1.7\pm0.15\;M_{\odot}$ -- see \cite{tlk11a} for a detailed description.
    }
\label{fig:IMXB4}
\end{center}
\end{figure*}

\subsection{Close IMXB systems -- Case~A RLO and a long-lasting, stable mass transfer}
The formation of MSPs from close-orbit IMXBs with main sequence donors has been studied in detail
by \citet{prp02,lrp+11,tlk11a,tl11}. The outcome of these systems is mainly MSPs with CO~WD companions (but
He~WD companions are also formed if the donor mass is at the low end or the initial binary
is very close). The MSPs formed through this channel are expected to be fully recycled (i.e. with spin periods of a few ms)
since the last part of the mass-transfer phase
lasts about $10^7\;{\rm yr}$ in this case (see Fig.~\ref{fig:IMXB4}). 
This is opposite to the preferentially mildly recycled MSPs formed by IMXBs via early Case~B or Case~C RLO (described above) where the duration of the
total accretion phase is only of
the order of $10^6\;{\rm yr}$, or even less for Case~C RLO which always evolves through a CE.

\subsubsection{PSR~J1614$-$2230}
To shown an example of an IMXB system of Case~A RLO we have shown in Fig.~\ref{fig:IMXB4} one possible formation path
for this recently discovered $1.97\;M_{\odot}$ MSP \citep{dpr+10}. It is interesting to notice that this MSP
is the only system with a CO~WD companion which is fully recycled ($P=3.15\;{\rm ms}$). However, this full recycling
of the pulsar is indeed expected if it evolved
from an IMXB of Case~A RLO, see above. 

\subsection{AIC -- accretion induced collapse (a half road to form a millisecond pulsar?)}
It has been suggested that a neutron star can form from the implosion of a massive ONeMg~WD
if such a WD is accreting material (within a certain rate) from a companion star
\citep[e.g.][and references therein]{mic87,cil90,nk91}. This companion can either be
a main sequence star (super-soft {X}-ray source), a low-mass giant (novae-like system) or a helium star.
The AIC route to form MSPs has two main advantages: 1) it may help to explain the postulated 
birthrate problem \citep{kn88} between the number of LMXB progenitor
systems and the observed number of MSPs, and 2) it can retain pulsars in globular clusters due
to the small momentum kick expected to be associated with the implosion.
(For details of simulations of the implosion mechanism, see e.g. \citet{dbo+06}). 
However, it seems difficult to predict the pulsar spin rate, as well as the surface magnetic field strength,
associated with a pulsar formed via AIC. On the other hand, one can argue that the newborn neutron star
formed via AIC might at a later stage -- once its donor star has recovered from the dynamical effects of sudden mass loss
(caused by the released gravitational binding energy in the transition from an accreting WD to a more compact neutron star) --
begin to accrete further material from its companion star, which should then resemble
the conditions under which 
MSPs are formed via the conventional channels.
A recent population synthesis study by \citet{htw+10} concludes that one cannot ignore the AIC route. However,
there are still uncertainties involved in such studies and in particular the applied conditions for
making the ONeMg~WD mass grow sufficiently. A weakness in advocating the AIC formation channel is
thus exactly the point that one cannot easily distinguish the result of this path from the standard scenario
(see, however, Section~\ref{sec:corbet} for a novel speculative hypothesis).
It would be very interesting though, if observations would yield either 
a very slowly spinning (few hundred ms) pulsar associated with a very low B-field, or
a high B-field MSP with a very rapid spin.
(However, the latter 
kind of pulsar would be very unlikely to be detected given that its very strong magnetic dipole radiation
would slow down its spin rate within a few Myr).

\begin{table*}
\center
\caption{Progenitor systems leading to the formation of MSPs -- see also Fig.~\ref{fig:cartoon}. 
         All the characteristic values for both the {X}-binaries and the MSP systems are only rough indications and
         depend on effects which are poorly known -- for example, concerning the strength of magnetic braking and other spin-orbit couplings,
         as well as the CE and spiral-in evolution. AICs are not included in this table.} 
\begin{tabular}{lcccl}
\noalign{\smallskip}
\hline
\noalign{\smallskip}
    {\bf LMXB}                                &  Case A                & Case B                   & & Comment\\
\hline
\noalign{\smallskip}
\hspace{0.3cm} Donor mass ($M_{\odot}$)       & $1-2$                  & $1-2 $                   & & \\
\hspace{0.3cm} $P_{\rm RLO}$ (days)           & $\le 1.0^{\ast}$       & $> 1.0^{\ast}$           & & $^{\ast}\,P_{\rm bif}$ uncertain\\
\hline
\noalign{\smallskip}
 \hspace{0.15cm} MSP                          & $\downarrow$           & $\downarrow$             & &  \\
\hline
\noalign{\smallskip}
\hspace{0.3cm} $P_{\rm orb}$ (days)           & $\le 1.0$              &    $1-1000$              & & \\
\hspace{0.3cm} Recycling                      & full                   &    full$^{\ast}$         & &  $^{\ast}$if $P_{\rm orb}\le 200$\\
\hspace{0.3cm} Companion                      & S$^{\ast\ast}$/UL/He   &    He                    & &  $^{\ast\ast}$S: single MSP\\
\hspace{0.3cm} Example PSR                    & B1957+20               &    1713+0747             & &  $\;\;$UL: ultra light\\
\hline
\noalign{\bigskip}
\noalign{\bigskip}
\hline
\noalign{\smallskip}
    {\bf IMXB}                                & Case A                 & early Case B             & Case~C           & Comment\\
\hline
\noalign{\smallskip}
\hspace{0.3cm} Donor mass ($M_{\odot}$)       & $3-5$                  & $3-6 $                   & $3-9$            & Case C $\rightarrow$ CE\\
\hspace{0.3cm} $P_{\rm RLO}$ (days)           & $\le 2.5$              & $3-10$                   & $100-1000$       & \\
\hline
\noalign{\smallskip}
 \hspace{0.15cm} MSP                          & $\downarrow$           & $\downarrow$             & $\downarrow$     &  \\
\hline
\noalign{\smallskip}
\hspace{0.3cm} $P_{\rm orb}$ (days)           & $3-20$                 &  $3-50$                  & $0-20^{\ast}$    & $^{\ast}\,\lambda,\eta$ uncertain\\
\hspace{0.3cm} Recycling                      & full                   &  partial                 & partial          & \\
\hspace{0.3cm} Companion                      & CO (He)                &    CO                    & CO/ONeMg         & WDs\\
\hspace{0.3cm} Example PSR                    & 1614$-$2230            &  0621+1002               & 1802$-$2124      & \\
\hline
\end{tabular}
\label{table:classes}
\end{table*}
\begin{landscape}
\begin{figure*}[h]
\begin{center}
\mbox{\includegraphics[width=1.30\textwidth, angle=0]{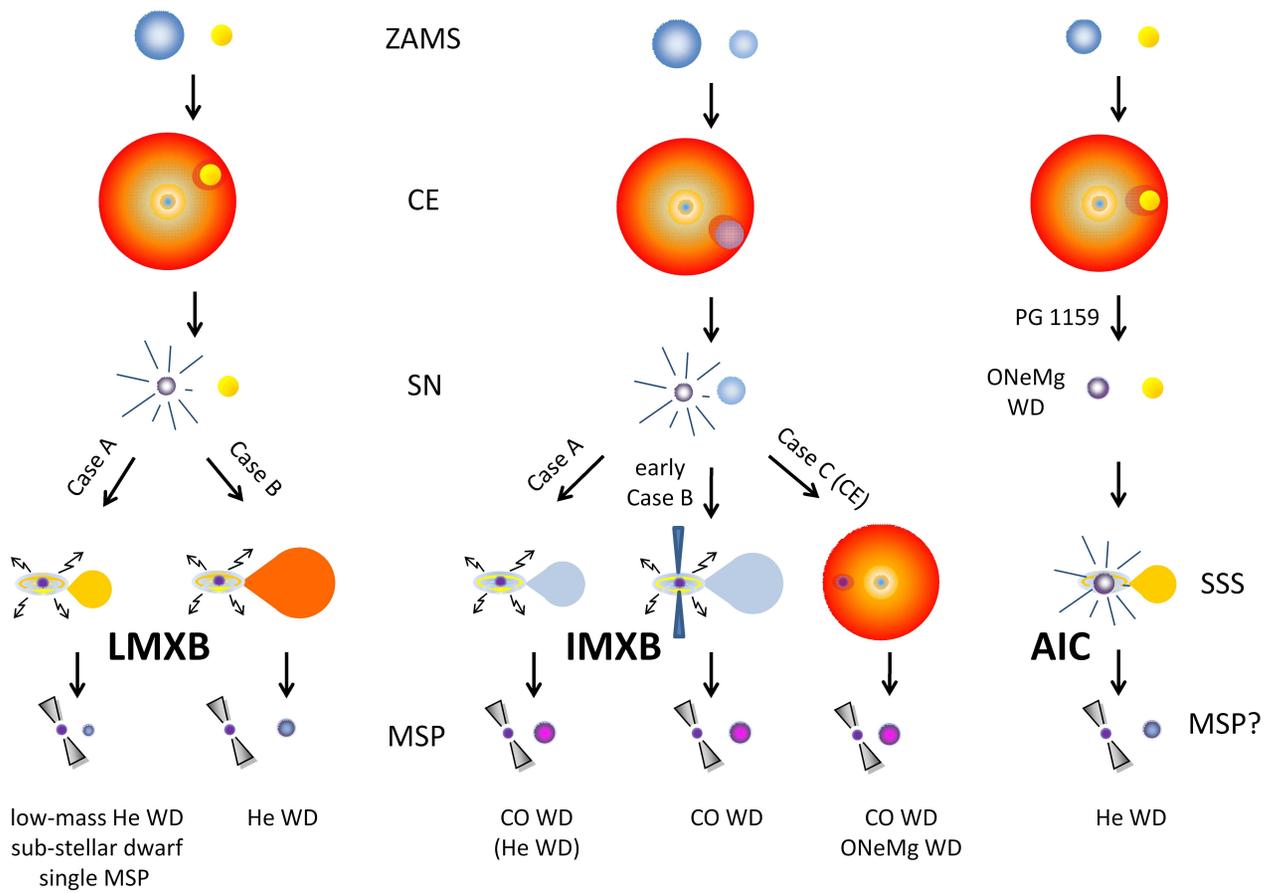}}
  \caption[]{
    Formation of MSPs. From left to right: LMXBs (Cases A and B), IMXBs (Cases A, early B and C (CE)) and possibly AIC?
    }
\label{fig:cartoon}
\end{center}
\end{figure*}
\end{landscape}

\section{The Corbet diagram revisited -- a new puzzle? (related to AIC?)}
\label{sec:corbet}
\begin{figure*}[h!]
\begin{center}
\mbox{\includegraphics[width=0.60\textwidth, angle=-90]{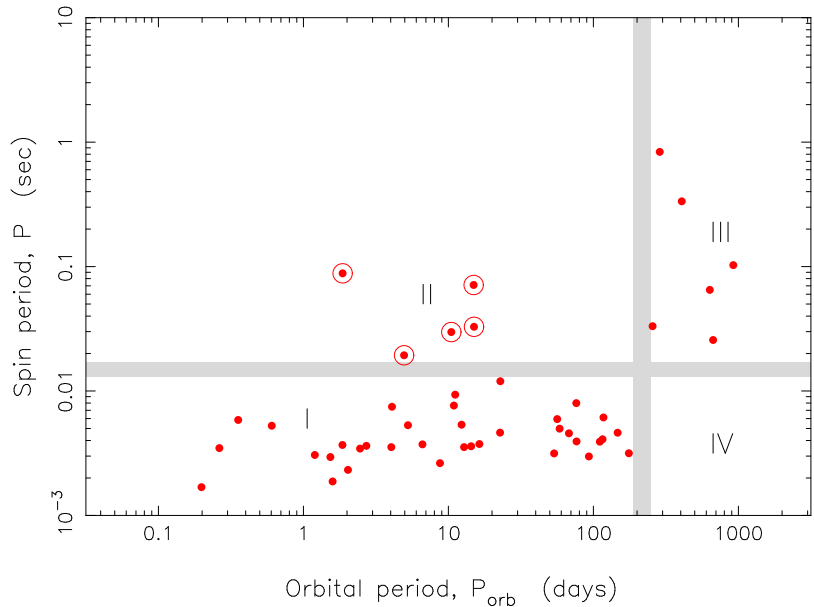}}
  \caption[]{
    The distribution of the 47 binary radio pulsars with a He~WD companion
    in the Corbet diagram. The plot reveals four regions marked by I, II, III and IV. 
    These regions may be understood from an evolutionary
    point of view -- see text..
    The pulsars in region~II are also marked by a ring in the $P\dot{P}$-diagram in Fig.~\ref{fig:ppdot}.
    }
\label{fig:corbet}
\end{center}
\end{figure*}

Plotting all binary pulsars in the Corbet diagram does not reveal much interesting information - it is a big mess!
However, if one only considers the population of pulsars with He~WD companions an interesting pattern is noticed,
as shown in Fig.~\ref{fig:corbet}. Region~I shows that MSPs can be fully recycled over a spread of 3 orders of magnitude in
final orbital period. (Note, the well-known orbital period gap in region~I, currently between 23 and 53~days).
If the orbital period exceeds $\sim\!200~{\rm days}$ the pulsars are only partially recycled
as noticed from their slow spin periods (region~III). This is expected since the mass-transfer phase is relatively short in those wide LMXBs where
the donor star is highly evolved by the time it fills its Roche-lobe \citep{ts99}.
This may explain why no pulsars are seen in region~IV.
In region~II one sees a sub-population of systems with orbital periods between 1 and 20~days, all of which are only mildly
recycled MSPs with spin periods between 20 and 100~ms. Where do these systems come from? Obviously from progenitor systems
where the mass transfer was limited. Whereas the populations in regions~I and III are produced from LMXBs with very 
different initial periods
the situation might be different for region~II.
Could these systems then perhaps originate from AICs where the subsequent spin~up of the newborn neutron star
only resulted in a mild spin-up? (because of the 
limited material remaining in the donor star envelope following the mass transfer to the ONeMg~WD before the implosion).
The narrow range of orbital periods for these pulsars could then
reflect the fine-tuned interval of allowed mass-transfer rates needed for the progenitor ONeMg~WD to accrete and grow in mass to the
Chandrasekhar limit before its implosion.

\section{Spinning up old neutron stars to millisecond pulsars}
\label{sec:spinup}
\begin{figure*}[t]
\begin{center}
\mbox{\includegraphics[width=0.50\textwidth, angle=-90]{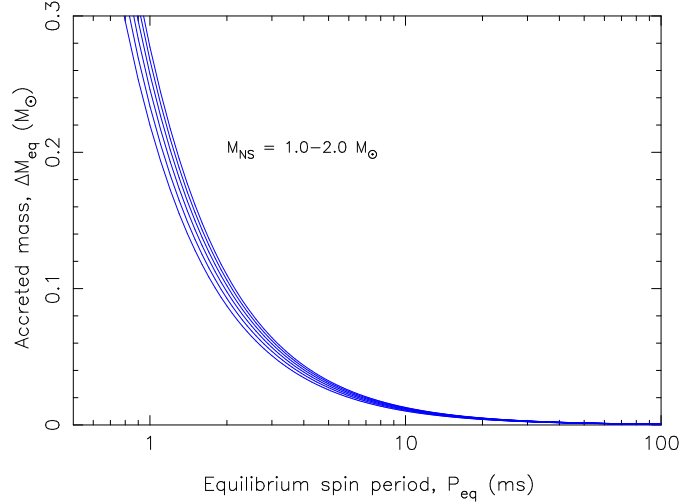}}
  \caption[]{
     The amount of mass needed to spin~up a pulsar as a function of its equilibrium spin period -- see text for discussion.
     The curves correspond to different neutron star masses between
     $1.0-2.0\;M_{\odot}$ (in steps of $0.2\,M_{\odot}$) increasing upwards.
    }
\label{fig:deltaM}
\end{center}
\end{figure*}

The physics of the accretion disk--magnetosphere interaction is still not known in detail.
The interplay between the neutron star magnetic field and the conducting plasma in the disk
is a rather complex process.
For details of the accretion physics I refer to \citet{pr72,lpp73,st83,gl92,st93,fkr02,rfs04,spi06} and references therein.
When the conditions for accretion are 
optimal the spin-up torque, $N=\partial J / \partial t$, acting at the edge of the co-rotating magnetosphere 
(roughly at the distance of the Alfv\'en radius, $r_A$) is easily estimated
from the specific orbital angular momentum at this location ($j=|\vec{r}\times \vec{v}|\simeq r_A^2\Omega=\sqrt{GMr_A}$) and is roughly:
\begin{equation} 
  N = \frac{\partial}{\partial t}(I\Omega)\sim \dot{M}\sqrt{GMr_A}\,f
\end{equation} 
where $M$ is the mass of the neutron star, $\dot{M}$ is its accretion rate and $f$ is a factor of order unity depending on the accretion flow. 
Observations of the spin evolution of accreting neutron stars at a given instant show rather stochastic variations, even with torque reversals
\citep{bcc+97}.
However, the long-term spin rate will eventually tend toward the equilibrium spin period, $P_{\rm eq}$.
The location of the associated so-called spin-up line for the rejuvenated pulsar in the $P\dot{P}$-diagram can be found by
considering the equilibrium configuration when the angular velocity of the neutron star is equal to the
Keplerian angular velocity of the magnetosphere, at roughly the Alfv\'en surface, where the accreted matter enters the magnetosphere,
i.e. $\Omega _\star = \Omega _{\rm eq} = \Omega _{\rm K}(r_{\rm A})$ or:
\begin{equation}
     P_{\rm eq} = 2\pi \sqrt{\frac{r_{\rm A}^3}{GM}}\,\frac{1}{\omega _c}
  \label{eq:Peq0}
\end{equation}
where $\omega _c$ is the so-called critical fastness parameter which
is a measure of where the accretion torque vanishes (depending on the dynamical importance
of the pulsar spin rate).
\citet{tlk11b} recently demonstrated
that one can obtain
a convenient expression for the amount of accreted mass needed to spin~up a pulsar:
\begin{equation}
     \Delta M_{\rm eq} \simeq 0.22\,M_{\odot}\; \frac{(M/M_{\odot})^{1/3}}{P_{\rm ms}^{4/3}}
  \label{eq:deltaMfinalfit}
\end{equation}
where $M$ is the mass of the recycled pulsar (after spin-up) and the equilibrium spin period is given in ms.
In Fig.~\ref{fig:deltaM} we show how much mass is needed to spin~up a pulsar to a given spin period.
For example, it requires 0.003, 0.01 and $0.10\;M_{\odot}$ to spin~up a pulsar to a spin period of 
25, 10 and $2~{\rm ms}$, respectively.
The value of $\Delta M_{\rm eq}$ should be regarded as a {\it lower} limit to the actual amount of material
transfered since the accretion efficiency is less than unity -- even at sub-Eddington levels, see Section~\ref{subsec:acc-eff}.

It is a fact that the B-field of the neutron star decreases during the recycling process. However,
it is still not understood if
this is caused by spin-down induced flux expulsion of the core proton fluxoids \citep{sbmt90},
or if the B-field is confined to the crustal regions and decays due to diffusion and Ohmic dissipation, 
as a result of a decreased electrical conductivity when heating effects set in from nuclear burning of the accreted material
\citep{gu94,kb97},
or if the field decay is simply caused by a diamagnetic screening by the accreted plasma 
-- see \citet{bha02} and references therein.
The decay of the crustal B-field is found by solving the induction equation:
\begin{equation}
\frac{\partial \vec B}{\partial t} = - \frac{c^2}{4 \pi} \vec \nabla \times
(\frac{1}{\sigma} \times \vec \nabla \times \vec B)
 + \vec \nabla \times (\vec V \times \vec B)     
\label{induction}
\end{equation}
where the electrical conductivity, $\sigma$ depends on the temperature of the crust 
(as well as the local density and lattice impurities).

\section{Masses of millisecond pulsars}
In order to weigh a pulsar it must be a member of a binary system.
The most precisely measured masses of pulsars are obtained 
via general relativistic effects. The related post-Keplerian parameters include periastron advance,
redshift/time~dilation, orbital period derivative and Shapiro delay \citep[e.g.][]{Wil09}.
Shapiro delays of radio signals from pulsars \citep[][]{sac+98} have the advantage
of being measurable also in low eccentricity systems if the orbital inclination is high
(i.e. close to $90\deg$).
This method yields the opportunity to
weigh both compact stars accurately.
So far, such measurements have revealed masses of recycled neutron stars between
$1.17\;M_{\odot}$ \citep{jsk+08} and $1.97\;M_{\odot}$ \citep{dpr+10}.
This wide spread of recycled pulsar masses reveals both the various degree of recycling and a spread in the birth masses of
these neutron stars -- see extended discussion in \citet{tlk11a}.

\subsection{An anti-correlation between pulsar mass and orbital period?}
For pulsars with He~WD companions it has been argued that
an anti-correlation between pulsar mass and orbital period would be expected as a simple consequence of the
interplay between mass-transfer rate (and thus accretion rate), orbital period and the
evolutionary status of the LMXB donor star at the onset of the RLO \citep[see][for details]{ts99}.
So far, this anti-correlation has not been confirmed from the few systems with measured pulsar masses.
Keep also in mind that this anti-correlation is probably blurred by the fact that neutron stars
are born with different masses.

\subsection{The accretion efficiency}
\label{subsec:acc-eff}
\citet{ts99} noticed from a comparison between observational constraints on pulsar masses and their calculated LMXB models
that the accretion efficiency is sometimes less than 50\% -- even for those LMXBs where the neutron star accretes at sub-Eddington rates.
The conclusion is that a significant amount of matter may be ejected from the pulsar magnetosphere due to magnetodipole wave pressure,
the Gunn-Ostriker mechanism
or the propeller effect \citep{is75}. Furthermore, accretion disk instabilities \citep{pri81,vpa96}
are also responsible for ejecting part of the transfered material.
Hopefully, future measurements of neutron star masses can help us to constrain the accretion efficiency better.

\section{Summary}
When trying to determine the progenitor system of a given observed binary MSP there are a number of important parameters to consider:
the pulsar spin period, its period derivative, the orbital period, the eccentricity and the masses of both the companion star and
the pulsar itself. Given these properties, it should be possible to place the MSP among the five (and a half)
formation channels summarized in this paper. (The origin of the double neutron star systems is not considered in this review, but
has an equally interesting story). Of course, there are constantly discovered new oddballs trying to destroy our general picture
of MSP formation, calling for a continuous revision of possibilities. For example, the discovery of the binary MSP~J1903+0327 \citep{fbw+11}
which requires a triple system origin to explain its current peculiar characteristics 
(fully recycled, high eccentricity, wide orbit, $1\;M_{\odot}$ main sequence companion star).

The most important parameters for forecasting the outcome of an {X}-ray binary system are the mass ratio of the stellar components
and the evolutionary status of the donor star at the onset of RLO.
The nature of the observed MSP companion star tells us if the progenitor systems was an LMXB or an IMXB.

The MSPs which are only mildly recycled must have had a limited mass-transfer phase: either because their  
companion star at this stage was relative massive and/or evolved enough to have a convective envelope, 
or it might be connected to an AIC event.
The Corbet diagram for these binary pulsars with He~WD companions seems to include important informations about the origin of these
mildly recycled MSPs.

It is of great theoretical desire to have measured as many pulsar and companion masses as possible. These values are important for understanding
neutron star formation, binary stellar evolution and accretion processes. 

It was my intention to describe the beautiful link between observations of MSPs and binary stellar astrophysics. 
If the reader finds this paper somewhat boring it is certainly not the fault of the topic.

\acknowledgements I am very grateful for the financial support and collaboration with Norbert~Langer and Michael~Kramer.
I also thank the LOC for inviting me to give this review. (It was a great conference in Chile).
Finally, many thanks to Joris Verbiest for giving detailed comments to the manuscript.
\bibliographystyle{asp2010}
\bibliography{pulsars}

\begin{thebibliography}{}
\expandafter\ifx\csname natexlab\endcsname\relax\def\natexlab#1{#1}\fi
\expandafter\ifx\csname url\endcsname\relax
  \def\url#1{\texttt{#1}}\fi
\expandafter\ifx\csname urlprefix\endcsname\relax\def\urlprefix{URL }\fi
\providecommand{\eprint}[2][]{\url{#2}}

\bibitem[{Alpar et~al.(1982)Alpar, Cheng, Ruderman, \& Shaham}]{acrs82}
Alpar, M.~A., Cheng, A.~F., Ruderman, M.~A., \& Shaham, J. 1982, \nat, 300, 728

\bibitem[{{Applegate} \& {Shaham}(1994)}]{as94}
{Applegate}, J.~H., \& {Shaham}, J. 1994, \apj, 436, 312

\bibitem[{{Archibald} et~al.(2009){Archibald}, Stairs, Ransom, Kaspi,
  Kondratiev, Lorimer, McLaughlin, Boyles, Hessels, Lynch, van Leeuwen,
  Roberts, Jenet, Champion, Rosen, Barlow, Dunlap, \& Remillard}]{asr+09}
{Archibald}, A.~M., Stairs, I.~H., Ransom, S.~M., Kaspi, V.~M., Kondratiev,
  V.~I., Lorimer, D.~R., McLaughlin, M.~A., Boyles, J., Hessels, J.~W.~T.,
  Lynch, R., van Leeuwen, J., Roberts, M.~S.~E., Jenet, F., Champion, D.~J.,
  Rosen, R., Barlow, B.~N., Dunlap, B.~H., \& Remillard, R.~A. 2009, Science,
  324, 1411

\bibitem[{{Backer} et~al.(1982){Backer}, {Kulkarni}, {Heiles}, {Davis}, \&
  {Goss}}]{bkh+82}
{Backer}, D.~C., {Kulkarni}, S.~R., {Heiles}, C., {Davis}, M.~M., \& {Goss},
  W.~M. 1982, \nat, 300, 615

\bibitem[{{Bhattacharya}(2002)}]{bha02}
{Bhattacharya}, D. 2002, Journal of Astrophysics and Astronomy, 23, 67

\bibitem[{Bhattacharya \& {van den Heuvel}(1991)}]{bv91}
Bhattacharya, D., \& {van den Heuvel}, E. P.~J. 1991, Physics Reports, 203, 1

\bibitem[{{Bildsten} et~al.(1997){Bildsten}, {Chakrabarty}, {Chiu}, {Finger},
  {Koh}, {Nelson}, {Prince}, {Rubin}, {Scott}, {Stollberg}, {Vaughan},
  {Wilson}, \& {Wilson}}]{bcc+97}
{Bildsten}, L., {Chakrabarty}, D., {Chiu}, J., {Finger}, M.~H., {Koh}, D.~T.,
  {Nelson}, R.~W., {Prince}, T.~A., {Rubin}, B.~C., {Scott}, D.~M.,
  {Stollberg}, M., {Vaughan}, B.~A., {Wilson}, C.~A., \& {Wilson}, R.~B. 1997,
  \apjs, 113, 367

\bibitem[{{Camilo}(1996)}]{cam96}
{Camilo}, F. 1996, in IAU Colloq. 160: Pulsars: Problems and Progress, edited
  by {S.~Johnston, M.~A.~Walker, \& M.~Bailes}, vol. 105 of Astronomical
  Society of the Pacific Conference Series, 539

\bibitem[{{Canal} et~al.(1990){Canal}, {Isern}, \& {Labay}}]{cil90}
{Canal}, R., {Isern}, J., \& {Labay}, J. 1990, \araa, 28, 183

\bibitem[{{Deloye} \& {Bildsten}(2003)}]{db03}
{Deloye}, C.~J., \& {Bildsten}, L. 2003, \apj, 598, 1217

\bibitem[{{Demorest} et~al.(2010){Demorest}, {Pennucci}, {Ransom}, {Roberts},
  \& {Hessels}}]{dpr+10}
{Demorest}, P.~B., {Pennucci}, T., {Ransom}, S.~M., {Roberts}, M.~S.~E., \&
  {Hessels}, J.~W.~T. 2010, \nat, 467, 1081

\bibitem[{{Dessart} et~al.(2006){Dessart}, {Burrows}, {Ott}, {Livne}, {Yoon},
  \& {Langer}}]{dbo+06}
{Dessart}, L., {Burrows}, A., {Ott}, C.~D., {Livne}, E., {Yoon}, S.-C., \&
  {Langer}, N. 2006, \apj, 644, 1063

\bibitem[{{Dewi} et~al.(2006){Dewi}, {Podsiadlowski}, \& {Sena}}]{dps06}
{Dewi}, J.~D.~M., {Podsiadlowski}, P., \& {Sena}, A. 2006, \mnras, 368, 1742

\bibitem[{{Dewi} \& {Pols}(2003)}]{dp03}
{Dewi}, J.~D.~M., \& {Pols}, O.~R. 2003, \mnras, 344, 629

\bibitem[{{Dewi} \& {Tauris}(2000)}]{dt00}
{Dewi}, J. D.~M., \& {Tauris}, T.~M. 2000, \aap, 360, 1043

\bibitem[{{Frank} et~al.(2002){Frank}, {King}, \& {Raine}}]{fkr02}
{Frank}, J., {King}, A., \& {Raine}, D.~J. 2002, {Accretion Power in
  Astrophysics: Third Edition} (Cambridge University Press)

\bibitem[{{Freire} et~al.(2011){Freire}, {Bassa}, {Wex}, {Stairs}, {Champion},
  {Ransom}, {Lazarus}, {Kaspi}, {Hessels}, {Kramer}, {Cordes}, {Verbiest},
  {Podsiadlowski}, {Nice}, {Deneva}, {Lorimer}, {Stappers}, {McLaughlin}, \&
  {Camilo}}]{fbw+11}
{Freire}, P.~C.~C., {Bassa}, C.~G., {Wex}, N., {Stairs}, I.~H., {Champion},
  D.~J., {Ransom}, S.~M., {Lazarus}, P., {Kaspi}, V.~M., {Hessels}, J.~W.~T.,
  {Kramer}, M., {Cordes}, J.~M., {Verbiest}, J.~P.~W., {Podsiadlowski}, P.,
  {Nice}, D.~J., {Deneva}, J.~S., {Lorimer}, D.~R., {Stappers}, B.~W.,
  {McLaughlin}, M.~A., \& {Camilo}, F. 2011, \mnras, 412, 2763

\bibitem[{{Geppert} \& {Urpin}(1994)}]{gu94}
{Geppert}, U., \& {Urpin}, V. 1994, \mnras, 271, 490

\bibitem[{{Ghosh} \& {Lamb}(1992)}]{gl92}
{Ghosh}, P., \& {Lamb}, F.~K. 1992, in X-Ray Binaries and the Formation of
  Binary and Millisecond Radio Pulsars (NATO Science Series), 487

\bibitem[{{Hurley} et~al.(2010){Hurley}, {Tout}, {Wickramasinghe}, {Ferrario},
  \& {Kiel}}]{htw+10}
{Hurley}, J.~R., {Tout}, C.~A., {Wickramasinghe}, D.~T., {Ferrario}, L., \&
  {Kiel}, P.~D. 2010, \mnras, 402, 1437

\bibitem[{{Illarionov} \& {Sunyaev}(1975)}]{is75}
{Illarionov}, A.~F., \& {Sunyaev}, R.~A. 1975, \aap, 39, 185

\bibitem[{{Janssen} et~al.(2008){Janssen}, {Stappers}, {Kramer}, {Nice},
  {Jessner}, {Cognard}, \& {Purver}}]{jsk+08}
{Janssen}, G.~H., {Stappers}, B.~W., {Kramer}, M., {Nice}, D.~J., {Jessner},
  A., {Cognard}, I., \& {Purver}, M.~B. 2008, \aap, 490, 753

\bibitem[{{Joss} et~al.(1987){Joss}, {Rappaport}, \& {Lewis}}]{jrl87}
{Joss}, P.~C., {Rappaport}, S., \& {Lewis}, W. 1987, \apj, 319, 180

\bibitem[{{Konar} \& {Bhattacharya}(1997)}]{kb97}
{Konar}, S., \& {Bhattacharya}, D. 1997, \mnras, 284, 311

\bibitem[{{Kulkarni} \& {Narayan}(1988)}]{kn88}
{Kulkarni}, S.~R., \& {Narayan}, R. 1988, \apj, 335, 755

\bibitem[{{Lamb} et~al.(1973){Lamb}, {Pethick}, \& {Pines}}]{lpp73}
{Lamb}, F.~K., {Pethick}, C.~J., \& {Pines}, D. 1973, \apj, 184, 271

\bibitem[{{Lazaridis} et~al.(2011){Lazaridis}, {Verbiest}, {Tauris},
  {Stappers}, {Kramer}, {Wex}, {Jessner}, {Cognard}, {Desvignes}, {Janssen},
  {Purver}, {Theureau}, {Bassa}, \& {Smits}}]{lvt+11}
{Lazaridis}, K., {Verbiest}, J.~P.~W., {Tauris}, T.~M., {Stappers}, B.~W.,
  {Kramer}, M., {Wex}, N., {Jessner}, A., {Cognard}, I., {Desvignes}, G.,
  {Janssen}, G.~H., {Purver}, M.~B., {Theureau}, G., {Bassa}, C.~G., \&
  {Smits}, R. 2011, \mnras, 584

\bibitem[{{Lin} et~al.(2011){Lin}, {Rappaport}, {Podsiadlowski}, {Nelson},
  {Paxton}, \& {Todorov}}]{lrp+11}
{Lin}, J., {Rappaport}, S., {Podsiadlowski}, P., {Nelson}, L., {Paxton}, B., \&
  {Todorov}, P. 2011, ArXiv e-prints, 1012.1877

\bibitem[{{Ma} \& {Li}(2009)}]{ml09}
{Ma}, B., \& {Li}, X.-D. 2009, \apj, 691, 1611

\bibitem[{{Manchester} et~al.(2005){Manchester}, {Hobbs}, {Teoh}, \&
  {Hobbs}}]{mhth05}
{Manchester}, R.~N., {Hobbs}, G.~B., {Teoh}, A., \& {Hobbs}, M. 2005, \aj, 129,
  1993

\bibitem[{{Michel}(1987)}]{mic87}
{Michel}, F.~C. 1987, \nat, 329, 310

\bibitem[{{Nagase}(1989)}]{nag89}
{Nagase}, F. 1989, \pasj, 41, 1

\bibitem[{{Nomoto} \& {Kondo}(1991)}]{nk91}
{Nomoto}, K., \& {Kondo}, Y. 1991, \apjl, 367, L19

\bibitem[{{Phinney}(1992)}]{phi92}
{Phinney}, E.~S. 1992, Royal Society of London Philosophical Transactions
  Series A, 341, 39

\bibitem[{{Podsiadlowski}(1991)}]{pod91}
{Podsiadlowski}, P. 1991, \nat, 350, 136

\bibitem[{{Podsiadlowski} et~al.(2002){Podsiadlowski}, {Rappaport}, \&
  {Pfahl}}]{prp02}
{Podsiadlowski}, P., {Rappaport}, S., \& {Pfahl}, E.~D. 2002, \apj, 565, 1107

\bibitem[{{Pringle}(1981)}]{pri81}
{Pringle}, J.~E. 1981, \araa, 19, 137

\bibitem[{{Pringle} \& {Rees}(1972)}]{pr72}
{Pringle}, J.~E., \& {Rees}, M.~J. 1972, \aap, 21, 1

\bibitem[{{Pylyser} \& {Savonije}(1988)}]{ps88}
{Pylyser}, E., \& {Savonije}, G.~J. 1988, \aap, 191, 57

\bibitem[{{Pylyser} \& {Savonije}(1989)}]{ps89}
{Pylyser}, E.~H.~P., \& {Savonije}, G.~J. 1989, \aap, 208, 52

\bibitem[{{Radhakrishnan} \& {Srinivasan}(1982)}]{rs82}
{Radhakrishnan}, V., \& {Srinivasan}, G. 1982, Current Science, 51, 1096

\bibitem[{{Ransom} et~al.(2005){Ransom}, {Hessels}, {Stairs}, {Freire},
  {Camilo}, {Kaspi}, \& {Kaplan}}]{rhs+05}
{Ransom}, S.~M., {Hessels}, J.~W.~T., {Stairs}, I.~H., {Freire}, P.~C.~C.,
  {Camilo}, F., {Kaspi}, V.~M., \& {Kaplan}, D.~L. 2005, Science, 307, 892

\bibitem[{{Rappaport} et~al.(1995){Rappaport}, {Podsiadlowski}, {Joss}, {Di
  Stefano}, \& {Han}}]{rpj+95}
{Rappaport}, S., {Podsiadlowski}, P., {Joss}, P.~C., {Di Stefano}, R., \&
  {Han}, Z. 1995, \mnras, 273, 731

\bibitem[{{Rappaport} et~al.(2004){Rappaport}, {Fregeau}, \& {Spruit}}]{rfs04}
{Rappaport}, S.~A., {Fregeau}, J.~M., \& {Spruit}, H. 2004, \apj, 606, 436

\bibitem[{{Refsdal} \& {Weigert}(1971)}]{rw71}
{Refsdal}, S., \& {Weigert}, A. 1971, \aap, 13, 367

\bibitem[{{Savonije}(1987)}]{sav87}
{Savonije}, G.~J. 1987, \nat, 325, 416

\bibitem[{{Shapiro} \& {Teukolsky}(1983)}]{st83}
{Shapiro}, S.~L., \& {Teukolsky}, S.~A. 1983, {Black holes, white dwarfs, and
  neutron stars: The physics of compact objects} (New York, Wiley-Interscience,
  1983, 663 p.)

\bibitem[{{Soberman} et~al.(1997){Soberman}, {Phinney}, \& {van den
  Heuvel}}]{spv97}
{Soberman}, G.~E., {Phinney}, E.~S., \& {van den Heuvel}, E.~P.~J. 1997, \aap,
  327, 620

\bibitem[{{Spitkovsky}(2006)}]{spi06}
{Spitkovsky}, A. 2006, \apjl, 648, L51

\bibitem[{{Spruit} \& {Taam}(1993)}]{st93}
{Spruit}, H.~C., \& {Taam}, R.~E. 1993, \apj, 402, 593

\bibitem[{{Srinivasan} et~al.(1990){Srinivasan}, {Bhattacharya}, {Muslimov}, \&
  {Tsygan}}]{sbmt90}
{Srinivasan}, G., {Bhattacharya}, D., {Muslimov}, A.~G., \& {Tsygan}, A.~J.
  1990, Current Science, 59, 31

\bibitem[{{Stairs} et~al.(1998){Stairs}, {Arzoumanian}, {Camilo}, {Lyne},
  {Nice}, {Taylor}, {Thorsett}, \& {Wolszczan}}]{sac+98}
{Stairs}, I.~H., {Arzoumanian}, Z., {Camilo}, F., {Lyne}, A.~G., {Nice}, D.~J.,
  {Taylor}, J.~H., {Thorsett}, S.~E., \& {Wolszczan}, A. 1998, \apj, 505, 352

\bibitem[{{Stairs} et~al.(2005){Stairs}, {Faulkner}, {Lyne}, {Kramer},
  {Lorimer}, {McLaughlin}, {Manchester}, {Hobbs}, {Camilo}, {Possenti},
  {Burgay}, {D'Amico}, {Freire}, \& {Gregory}}]{sfl+05}
{Stairs}, I.~H., {Faulkner}, A.~J., {Lyne}, A.~G., {Kramer}, M., {Lorimer},
  D.~R., {McLaughlin}, M.~A., {Manchester}, R.~N., {Hobbs}, G.~B., {Camilo},
  F., {Possenti}, A., {Burgay}, M., {D'Amico}, N., {Freire}, P.~C., \&
  {Gregory}, P.~C. 2005, \apj, 632, 1060

\bibitem[{{Tauris}(1996)}]{tau96}
{Tauris}, T.~M. 1996, \aap, 315, 453

\bibitem[{{Tauris} \& {Langer}(2011)}]{tl11}
{Tauris}, T.~M., \& {Langer}, N. 2011, in preparation

\bibitem[{{Tauris} et~al.(2011{\natexlab{a}}){Tauris}, {Langer}, \&
  {Kramer}}]{tlk11a}
{Tauris}, T.~M., {Langer}, N., \& {Kramer}, M. 2011{\natexlab{a}}, MNRAS,
  submitted, arXiv:astro-ph/1103.4996

\bibitem[{{Tauris} et~al.(2011{\natexlab{b}}){Tauris}, {Langer}, \&
  {Kramer}}]{tlk11b}
--- 2011{\natexlab{b}}, in preparation

\bibitem[{Tauris \& Savonije(1999)}]{ts99}
Tauris, T.~M., \& Savonije, G.~J. 1999, \aap, 350, 928

\bibitem[{{Tauris} \& {van den Heuvel}(2006)}]{tv06}
{Tauris}, T.~M., \& {van den Heuvel}, E.~P.~J. 2006, {Formation and evolution
  of compact stellar X-ray sources} (Cambridge University Press), chap.~16, 623

\bibitem[{{Tauris} et~al.(2000){Tauris}, {van den Heuvel}, \&
  {Savonije}}]{tvs00}
{Tauris}, T.~M., {van den Heuvel}, E.~P.~J., \& {Savonije}, G.~J. 2000, \apjl,
  530, L93

\bibitem[{{Tavani} \& {Brookshaw}(1992)}]{tav92}
{Tavani}, M., \& {Brookshaw}, L. 1992, \nat, 356, 320

\bibitem[{van~den Heuvel(1994)}]{vdh94a}
van~den Heuvel, E. P.~J. 1994, in Interacting Binaries, edited by
  H.~Nussbaumer, \& A.~Orr (Berlin: Springer-Verlag), 263

\bibitem[{{van den Heuvel}(1994)}]{vdh94b}
{van den Heuvel}, E.~P.~J. 1994, \aap, 291, L39

\bibitem[{{van der Sluys} et~al.(2005){van der Sluys}, {Verbunt}, \&
  {Pols}}]{vvp05}
{van der Sluys}, M.~V., {Verbunt}, F., \& {Pols}, O.~R. 2005, \aap, 440, 973

\bibitem[{{van Kerkwijk} et~al.(2005){van Kerkwijk}, {Bassa}, {Jacoby}, \&
  {Jonker}}]{vbjj05}
{van Kerkwijk}, M.~H., {Bassa}, C.~G., {Jacoby}, B.~A., \& {Jonker}, P.~G.
  2005, in Binary Radio Pulsars, edited by {F.~A.~Rasio \& I.~H.~Stairs}, vol.
  328 of Astronomical Society of the Pacific Conference Series, 357

\bibitem[{{van Paradijs}(1996)}]{vpa96}
{van Paradijs}, J. 1996, \apjl, 464, L139+

\bibitem[{{Voss} \& {Tauris}(2003)}]{vt03}
{Voss}, R., \& {Tauris}, T.~M. 2003, \mnras, 342, 1169

\bibitem[{{Webbink} et~al.(1983){Webbink}, {Rappaport}, \& {Savonije}}]{wrs83}
{Webbink}, R.~F., {Rappaport}, S., \& {Savonije}, G.~J. 1983, \apj, 270, 678

\bibitem[{Wijnands \& van~der Klis(1998)}]{wv98}
Wijnands, R., \& van~der Klis, M. 1998, \nat, 394, 344

\bibitem[{{Will}(2009)}]{Wil09}
{Will}, C.~M. 2009, \ssr, 148, 3

\bibitem[{{Wolszczan} \& {Frail}(1992)}]{wf92}
{Wolszczan}, A., \& {Frail}, D.~A. 1992, \nat, 355, 145

\end{thebibliography}

\end{document}